\documentclass[
     ,final            
   ]
   {aipproc}

\layoutstyle{6x9}

\begin{document}

\title{Exoplanets Around G--K Giants}

\classification{90}
\keywords {star: general - stars: fundamental parameters - stars: variable 
- techniques: radial velocities - stars: late-type - planetary systems}

\author{Michaela P.\ D\"{o}llinger}{
   address={Max-Planck-Institut f\"{u}r Astronomie, K\"{o}nigsstuhl 17, 
69117
Heidelberg, Germany}
   ,altaddress={European Southern Observatory, Karl-Schwarzschild-Stra\ss e 
2,
85748 Garching bei M\"{u}nchen, Germany} 
}
\author{Artie P.\ Hatzes}{
   address={Th\"{u}ringer Landessternwarte Tautenburg,
                 Sternwarte 5, 07778 Tautenburg, Germany}
}

\author{Luca Pasquini}{
   address={European Southern Observatory, Karl-Schwarzschild-Stra\ss e 2,
85748 Garching bei M\"{u}nchen, Germany}
}

\author{Eike~W.~Guenther}{
   address={Th\"{u}ringer Landessternwarte Tautenburg,
                 Sternwarte 5, 07778 Tautenburg, Germany}
}

\author{Michael Hartmann}{
   address={Th\"{u}ringer Landessternwarte Tautenburg,
                 Sternwarte 5, 07778 Tautenburg, Germany}
}

\author{Johny Setiawan}{
   address={Max-Planck-Institut f\"{u}r Astronomie, K\"{o}nigsstuhl 17, 
69117
Heidelberg, Germany}
}

\author{L\'eo Girardi}{
   address={INAF - Osservatorio Astronomico di Padova, Vicolo 
dell'Osservatorio 5, 35122, Padova, Italy}
}

\author{Jose R.\ de Medeiros}{
   address={Departamento de F{\'\i}sica,
     Universidade Federal do Rio Grande do Norte,
     Caixa Postal 1524, 59072-970, Natal, RN, Brasil}
}

\author{Licio da Silva}{
  address={Observat{\'o}rio Nacional - MCT,
    R.\ Gal.\ Jos{\'e} Cristino 77,
    20921-400, S{\~a}o Crist{\'o}v{\~a}o,
    Rio~de~Janeiro, RJ, Brasil}
}

\begin{abstract}
G and K giants are a class of radial velocity (RV) variables. One reason 
for
this variability are planetary companions which are indicated in time 
series
of stellar spectra. Since 2004 these spectra in the visual range were 
obtained
with the high resolution coud{\'e} {\'e}chelle spectrograph mounted on the
2m~telescope of the Th\"{u}ringer Landessternwarte Tautenburg 
(\textit{TLS}) for a
northern sample of 62~very bright K giants. In the South around 300~G and
K giants were observed with \textit{HARPS} mounted on the 3.6m~telescope 
on La Silla.
The \textit{TLS} sample contains at least 11~stars (18~$\%$) which show 
low-amplitude,
long-period RV variations most likely due to planets. This percentage of
planet frequency is confirmed by preliminary results of the \textit{HARPS} 
study.
Moreover the \textit{TLS} survey seems to indicate that giant planets do 
not favour
metal-rich stars, are more massive, and have longer periods than those 
found
around solar-type host stars.
\end{abstract}

\maketitle


\section{Introduction}
To date there are almost 500 extrasolar planets known orbiting other stars 
than our
Sun, mostly discovered via the RV method. In spite of this large effort,
current surveys are giving us a biased view of the process of planet 
formation
because less than 10~$\%$ of these planets orbit host stars with
masses~$M$~$>$~1.3~M$_{\odot}$.\par
Thus our knowledge about planet formation as a function of the most 
important
stellar parameter -- the mass of the host star -- is poorly understood. 
To make progress, the search for planets over a wider range of stellar 
masses
is essential. One strategy is to choose massive stars that have evolved 
off the
main sequence (MS). Thus giants have cool temperatures (more lines) and 
slow
rotation rates (narrow lines) resulting in excellent RV measurement 
precision.
G--K giants are a class of variable stars with multi-periodic RV 
variations
with different amplitudes and on two time scales.\par
The fact that the RV variability in giants has higher amplitudes than that
commonly seen in dwarfs suggests that it results from some specific
characteristics of these stars such as lower surface gravity.\par
The short-period (2--10 days) variations are likely caused by p-mode
oscillations.\par
In contrast the long-period ones occur on times scales of several hundreds 
of
days and can be due to orbiting stellar and sub-stellar companions as well 
as
rotational modulation due to star spots.\par
Doppler shifts caused by low-mass companions are expected to be extremely
stable with time. In addition they should not induce any variations in the
spectral line profile or be accompanied by variations in stellar activity
indicators.
We thus do not expect any correlation between the radial velocity 
behaviour
and bisector shape or the lines of the chromospheric activity indicators
Ca H $\&$ K.\par
By using the projected rotational velocity of the stars and their
radius, we can also check that the orbital period differs substantially 
from
the orbital period which will enable us to exclude rotational modulation.
If a large surface inhomogeneity passes the line-of-sight
of the observer as the star rotates, then this causes distortions in the
spectral line profiles that will be detected as RV variations with the
rotation period of the star.\par
Hatzes $\&$ Cochran (1993) found first indications of sub-stellar 
companions
around giants. The first extrasolar planet around the K~giant
HD~137759~($\iota$~Dra) was discovered by Frink et al.\ (2002). Other
exoplanets around HD~13189 and $\beta$~Gem were detected by
Hatzes et al.~(2005, 2006). The last planet was independently announced by
Reffert at al.\ (2006). Starting in 1998, Setiawan et al.\ (2003a) began 
to
search for planets around 83~giants with \textit{FEROS}. Up to now, this 
programme
has detected two giant exoplanets around HD~47536 (Setiawan et al.\ 
2003b), one
around HD~11977 (Setiawan et al.\ 2005), and more recently one around 
HD~110014
(de Medeiros et al.\ 2009). D\"ollinger (2008) started a similar survey in
February~2004 in the northern hemisphere monitoring a sample of
62~K~giant~stars using higher RV accuracy at \textit{TLS}. During this 
survey planets
around the K~giants 4~UMa (D\"ollinger et al.\ 2007), 42~Dra and HD~139357
(D\"ollinger 2009a), as well as 11~UMi and HD~32518 (D\"ollinger 2009b), 
which
most likely host extrasolar planets in almost circular orbits, were 
detected.\par
Moreover several surveys are actively searching for planets around giant
stars.\par
In 2001, Sato started a precise Doppler survey of about 300~G--K
giants (Sato et al.\ 2005) using a 1.88~m telescope at Okayama 
Astrophysical
Observatory (OAO). From this survey, planetary companions around HD~104985
(Sato et al.\ 2003), the Hyades giant $\epsilon$~Tau (Sato et al.\ 2007),
18~Del, $\xi$~Aql, and HD~81688 (Sato et al.\ 2008) were detected. 
Furthermore,
this survey discovered planetary companions around 14~And and 81~Cet
(Sato et al.\ 2008). In the same paper the detection of exoplanets
orbiting the subgiants 6~Lyn and HD~167042 were reported. Niedzielski et 
al.
(2007) discovered an exoplanet to the K0~giant HD~17092 using observations
taken with the Hobby-Eberly Telescope (\textit{HET}) between 2004~January 
and
2007~March. Johnson et al.\ (2007) published exoplanets around the three
intermediate-mass subgiants HD~192699, HD~210702, and HD~175541. Planetary
companions around two other subgiants HD~167042 and HD~142091 were 
discovered
monitoring a sample of 159 evolved stars at Lick and Keck Observatories 
for
the past 3.5~years by Johnson et al.\ (2008).
Liu et al.\ (2009) detected a planetary companion around the 
intermediate-mass
G~giant HD~173416.
Recently Sato et al.\ (2010) published a further planetary companion 
around
the K0~giant HD~145457. The detection of other seven exoplanets was 
recently
announced by Johnson et al.\ (2010).

\section{Data analysis}
Our spectra in the North have been taken since February 2004 by using the 
high
resolution spectrometer plus an iodine absorption cell mounted at the
Alfred-Jensch 2m~telescope in Tautenburg. 
This is a grism crossed-dispersed {\'e}chelle spectrograph that has a
resolution of $R(\frac{\lambda}{\delta \lambda}) \sim$ 67{,}000 and a
wavelength coverage of 4630--7370~$\mathring{A}$ when using the so-called
``visual'' (VIS) grism.\par
The high resolution and the large spectral range were required for the
determination of RVs and chemical abundances (D\"ollinger 2008).
A high resolving power was essential to guarantee a good wavelength
separation, which means that wavelengths with a small wavelength 
separation
of $\delta \lambda$ can be resolved. A large wavelength coverage was
necessary to achieve more accurate Doppler shift measurements by using 
more
spectral lines for radial velocity determinations and to provide enough Fe
lines for abundance analysis.\par
Both criteria are best achieved by cross-dispersed {\'e}chelle 
spectrographs
which use two separate dispersing elements. The spectral resolution is
reached with an {\'e}chelle grating used in high orders. A second
low-dispersion element such as a grism with an orthogonal dispersion axis
guarantees that the overlapping orders of the main grating do not fall on 
the
same pixels on the detector. This grating is a cross-disperser which 
produces
a full {\'e}chellogram on the detector.\par
The iodine absorption cell used was placed in the optical path in
front of the spectrograph slit. The resulting iodine absorption spectrum 
is
superposed on top of the stellar spectrum, providing a stable wavelength
reference against which the stellar RV is measured. 
For the data reduction (bias subtraction, flat-fielding, and extraction)
\textit{IRAF}~routines have been used.\par
RVs have been calculated by modeling the observed spectra with a high
signal-to-noise ratio (S/N) template of the star (without iodine) and a 
scan
of our iodine cell taken at a very high resolution of 300{,}000 with the
Fourier-Transform Spectrometer (FTS) of the McMath-Pierce telescope at 
Kitt
Peak National Observatory.\par
We compute the relative velocity shift between stellar and iodine 
absorption
lines as well as model the temporal and spatial variations of the 
instrument
profile. The spectrum is split up in typically 125 chunks, where the RV 
values
were determined for each chunk. The achieved RV accuracy is 
3--5~m\,s$^{-1}$.\par
To complete our global search we started in November 2006 a RV survey with 
the
\textit{HARPS} spectrograph mounted on the 3.6m~telescope at La Silla 
including the
best planet candidates from the previous \textit{FEROS} study (Setiawan 
2004). 
This spectrograph has a very high resolution of 100{,}000 resulting in a 
RV
accuracy of 1~m\,s$^{-1}$ using the simultaneous ThAr calibration method. 
The
spectral line shape analysis (bisector) is crucial for the confirmation of
planet candidates.

\section{Star samples}
The Tautenburg star sample consists of 62~giants covering the spectral 
types
K0--K5. 
The \textit{TLS} target stars are well distributed over the sky in right
ascension. Most of the sample stars have declinations greater than
+~45$^\circ$ which are circumpolar at the Th\"{u}ringer Landessternwarte
Tautenburg and so visibility over most of the year is guaranteed. 
In addition the stars are very bright which ensures short integration 
times. 
Their \textit{HIPPARCOS} parallaxes have an error of less than 10~$\%$ 
which was
essential for the determination of precise stellar parameters such as 
mass,
radius and age for each sample star (D\"ollinger 2008). 
Previously known binaries were excluded.
\par
The \textit{HARPS} sample contains 300~G--K giants including the 
\textit{FEROS} sample. 
About 80~G and K giant stars, with accurate \textit{HIPPARCOS} parallaxes, 
have been
systematically observed from October~1999 until February~2002 with 
\textit{FEROS}
attached at the 1.5m~ESO~telescope in La Silla.
\par
The new \textit{HARPS} targets were selected from the Bright Star (BS) 
catalogue. 
They are also bright and well distributed over the sky in right ascension. 
Giants with \textit{HIPPARCOS} parallaxes determined to be better than 10 
$\%$ and
having an intermediate (above 1.5 M$_{\odot}$) mass have been chosen. 
The last criteria is quite important to guarantee an enlarged mass range 
in
order to study planet formation as a function of the mass of the host 
star.

\section{Results}
The first preliminary results of these northern and southern surveys are 
now
available and will be presented in the next two subsections.

\subsection{Statistic of the Tautenburg survey}
\begin{itemize}
\item 2 stars (3 \%) are constant.
\item 15 stars (24 \%) exhibit short-period RV variations due to 
oscillations.
\item 12 stars (19 \%) belong to binary systems.
\item 11 stars (18 \%) exhibit long-term RV variations which are most 
likely caused by
   planetary companions.
\end{itemize}
%
A first result, after approximately 6 years of monitoring the giant 
\textit{TLS}
sample, is the detection of only 2 stars (3 $\%$) which show RV variations
lower than 10 m\,s$^{-1}$ at all time scales analyzed. It confirms that
K~giants are indeed RV variables. However, the term ``constant'' is
relative and very subjective. It depends on the measurement error and the
behaviour of the sample. Looking with enough precision and sampling there 
are
probably no ``constant'' K giant stars.\par
Moreover the statistics of the programme contain 15 stars (24 $\%$) which
show short-period RV variations possibly due to radial and/or non-radial
stellar oscillations.\par
In the case of stellar companions the most important discrimination 
criteria
are the comparatively very large RV amplitudes in the range of 
km\,s$^{-1}$ and
the long periods of more than several hundreds of days. This is due to the 
fact
that only the high masses of stellar companions, and not from planets, can
cause these high amplitudes. The large period as the second criterion 
excludes
furthermore short-period RV variations due to stellar oscillations. The 
third
criterion, the turnaround points of the orbit, is not visible in all RV 
curves.
This is a consequence of the large periods. It was only possible to
obtain a small part of the corresponding long orbits during our 
time-restricted
observations. Thus the RV curves of a part of the binary candidates show 
only a
snapshot of the whole orbit, expressed by the linear RV changes. The final
proof for a binary system is the calculation of an orbit. However, to
calculate such an orbit successfully, at least one turnaround point should 
be
visible in the RV curve and enough data points should be available.
At the moment for all 12~stars~(19~$\%$) of the \textit{TLS} sample the 
most likely
reason for the RV variations is that the stars belong to binary systems.
The percentage of the binaries is in very good agreement with results 
derived
from previous studies (e.g.\ Setiawan 2003a).\par
However, the most important result of the \textit{TLS} survey is the 
detection of
11~stars~(18~$\%$) which show low-amplitude, long-term radial velocity
variations on time scales of a few hundreds of days most likely due to
planetary companions.\\
Figure~1 shows the $\sigma_{RV}$ versus M$_{\mathrm{V}}$ diagram for the 
$TLS$ stars. In this plot the different types of RV variability are marked 
with different symbols. In general the distribution of the stars 
with planets is quite dispersed over the entire diagram which excludes most 
likely the presence of possible selection effects. 
 
\begin{figure}
\includegraphics[height=.7\textheight]{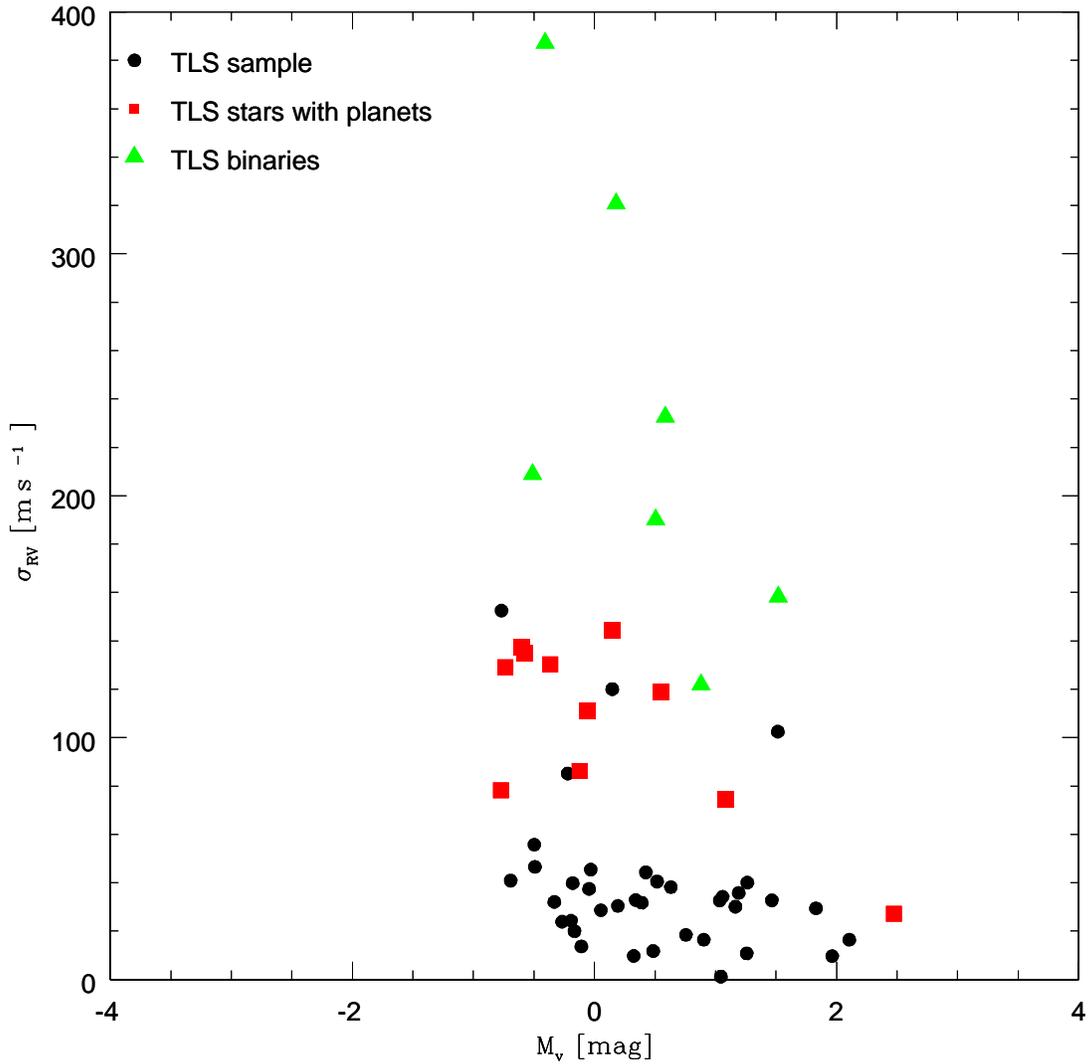}
\caption{RV variability versus M$_{\mathrm{V}}$ for the $TLS$ stars. In addition
 to a main body of stars showing increasing variability with luminosity, about 
20 stars stand out of this general trend, showing large RV variations. For most 
of them the cause of variability has been identified: planets (squares), 
brown dwarfs or stellar companions (triangles).}
\end{figure}

\subsection{Statistic of the \textit{FEROS} survey}

\begin{itemize}
\item 6 stars (8 \%) are constant.
\item 7 stars (9 \%) exhibit short-period RV variations due to
oscillations.
\item 15 stars (19 \%) belong to binary systems.
\item 5 stars (7 \%) show long-term RV variations which are most
likely induced by
   rotational modulation.
\item 7 stars (10 \%) exhibit long-term RV variations which are most
likely caused by
   planetary companions.
\end{itemize}
%
The southern sample has been monitored from October~1999 until February~2002 
with the $FEROS$ spectrograph attached at the 1.5m~ESO~telescope in La 
Silla. The first results of this southern survey have been published in
Setiawan et al. (2003a,b; 2004; 2005a,b; 2006), da Silva et al. (2006), 
Pasquini et al. (2007) and de Medeiros et al. (2009).\\
After $FEROS$ has been moved to the 2.2m~MPG/ESO~telescope, we have continued 
to follow-up a subsample of several giants showing clear evidence of 
long-period RV variations. The procedures adopted to compute the RV are 
explained in detail in Setiawan et al. (2003a).\\
Figure~2 shows the $\sigma_{RV}$ versus M$_{\mathrm{V}}$ diagram for the $FEROS$
stars. The general behavior is similar to what observed in Fig.~1 for the
$TLS$ sample. However, a direct comparison with the corresponding $TLS$ plot
(see Fig.~1) clearly shows that the variability observed in the $FEROS$ sample
is substantially higher than in the $TLS$ one.
We interpret this shift as due to the different levels of precision which were
obtained in both surveys: the $FEROS$ survey is less precise than the $TLS$
one.
Thanks to these follow-up observations we have detected 15~binary systems 
(19~$\%$).\\
For 5~target stars the long-period RV variations can be explained with
rotational modulation, because the RV periodical variability is related 
to the bisector and Ca~II variability. 
The $FEROS$ spectra cover a wavelength region of 360--920~nm with a typical
signal-to-noise ratio (SNR) of $\sim$~100--150. Thus, one can analyze the
stellar activity indicators like Ca~II~H and K, which are located at the
$\lambda$ values 393.4 and 396.7~nm, the Ca~II IR-triplet with the
corresponding $\lambda$ values of 894.8, 854.2 and 866.2~nm, as well as the
line profile asymmetry (bisector).
It is relevant to note as the majority of the stars showing rotational
modulation (see Fig.~2) have quite high luminosity, higher than most stars
in the $TLS$ sample. This is the reason why we do not find evidence for
these objects in the North.

\begin{figure}
\includegraphics[height=.7\textheight]{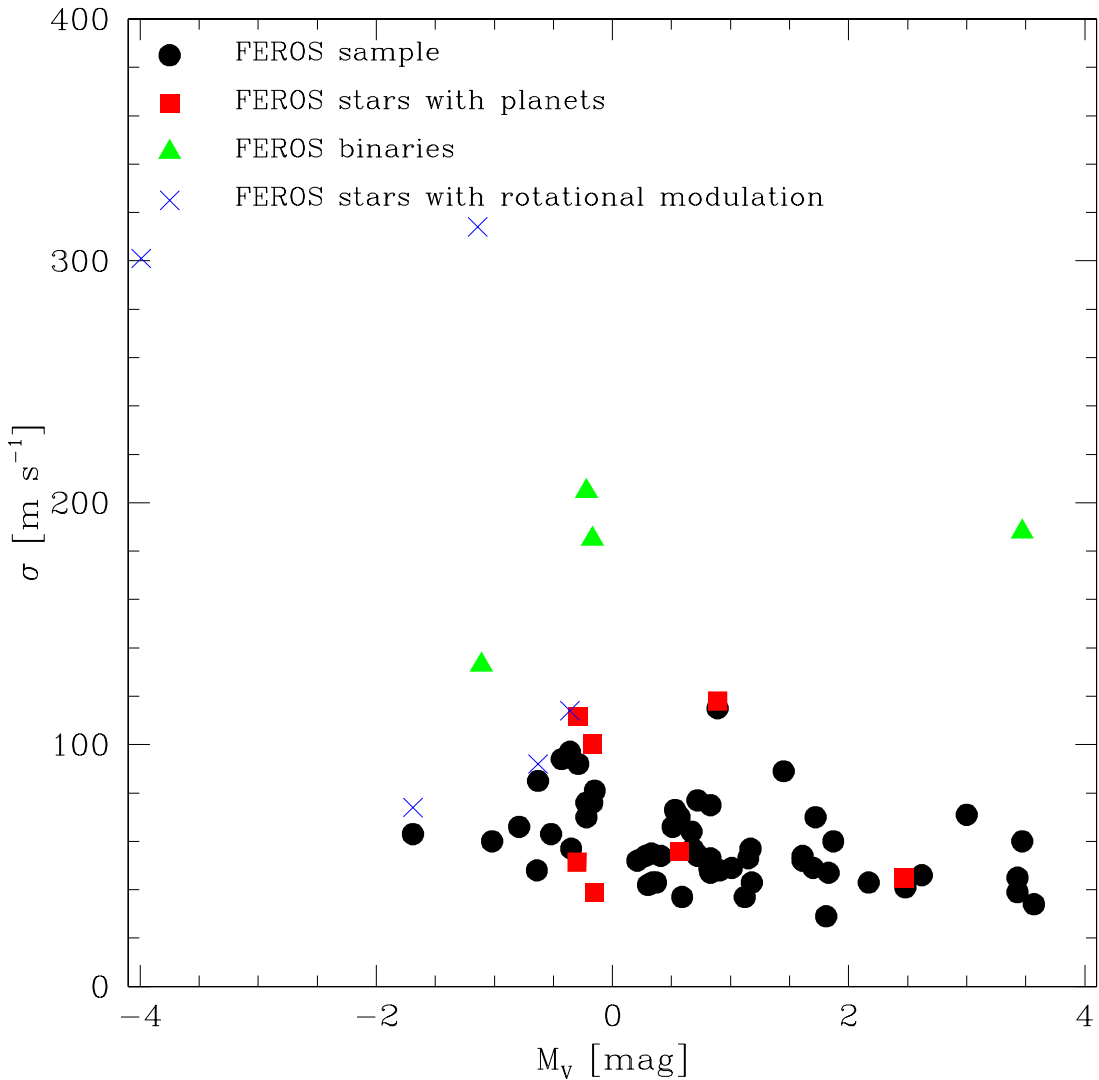}
\caption{RV variability versus M$_{\mathrm{V}}$ for the $FEROS$ sample. 
Symbols are as in Fig.~1. 
Crosses indicate stars for which RV variability is most 
likely induced by rotational modulation.}
\end{figure}

\subsection{Preliminary results from the \textit{TLS}, \textit{FEROS}, 
\textit{HARPS} and
other studies:}

Combining the northern and the southern star samples we are able to
summarize first results.

\begin{itemize}
\item Giant planets around giants are fairly common (about 10--15~$\%$). 
This percentage is higher than the frequency of $\approx$ 5 $\%$ for 
solar-type MS stars.\par
\item Planets around giant stars do not favour metal-rich stars (Pasquini
et al.\ 2007; Hekker $\&$ Melendez 2007; Hekker et al.\ 2008; Takeda et 
al.\
2008). A spectral analysis of the Tautenburg sample also confirms this 
behaviour (D\"ollinger 2008). This is in contrast to planet-hosting 
solar-type MS stars which tend to be metal-rich (e.g.\ Santos et al.\ 
2004).\par
\item Planets around giants have periods larger than $\sim$ 150~days.\par
\item Inner planets with orbital semi-major axes $a$~$\leq$~0.7~AU are not
present (Johnson et al.\ 2007; Sato et al.\ 2008).\par
\item Planets around giant stars have large masses, in the range of 
3--10~M$_{\mathrm{Jup}}$. For solar-type MS stars over half of the planets 
have masses less than 3~M$_{\mathrm{Jup}}$. For giant stars (intermediate 
stellar mass) over half of the planets have masses more than 
3--5~M$_{\mathrm{Jup}}$.
\end{itemize}

\section{Conclusions}
Although K giant variability can be complicated and can result from at 
least
three mechanisms, the future aim of our surveys is to distinguish between
these mechanisms and to verify the frequency of planets around G and K
giants for our samples.\par
Of course some of the above results are caused by observational or 
physical
biases; the detection of low-mass planets is for instance partially 
hampered
by the high intrinsic variability of giants (Setiawan et al.\ 2004, Hekker 
et
al.\ 2008), and the absence of short-period planets is naturally expected
because giants have large radii and they thus would have swallowed-up any
close-in planets. This region of the planetary orbital parameter space
($P$~$<$~20~days) is thus inaccessible.\par
In short, the existing results for planets around giant stars show a 
number of
properties which are different of those found around solar-type (and 
presumably
less massive) main-sequence stars.\par
Combining our northern and southern star sample we were able to put some 
of
these preliminary conclusions on firmer ground.

\bibliographystyle{aipproc}   
{}
\end{document}